\documentclass[a4paper]{article}

\usepackage{INTERSPEECH2021}

\title{
Robustifying automatic speech recognition by extracting slowly varying features 
\thanks{Accepted for the ISCA Symposium on Security and Privacy in Speech Communication (SPSC 2021).}
}
\name{Mat{\'\i}as Pizarro, 
Dorothea Kolossa, 
Asja Fischer 
}
\address{
  Ruhr University Bochum, Germany
  }
\email{\{Matias.PizarroBustamante, dorothea.kolossa, asja.fischer\}@rub.de}

\usepackage{xcolor}
\usepackage[normalem]{ulem}
\usepackage{multicol}
\usepackage{graphicx}
\usepackage{tabulary}
\usepackage{url}

\begin{document}

\maketitle
\begin{abstract}
  
    In the past few years, it has been shown that deep learning systems are highly vulnerable under attacks with adversarial examples.  Neural-network-based automatic speech recognition (ASR) systems are no exception. Targeted and untargeted attacks can modify an audio input signal in such a way that humans still recognise the same words, while ASR systems are steered to predict a different transcription.
    In this paper, we propose a defense mechanism against targeted adversarial attacks consisting in removing fast-changing features from the audio signals, either by applying slow feature analysis, a low-pass filter, or both, before feeding the input to the ASR system. We perform an empirical analysis of hybrid ASR models trained on data pre-processed in such a way. While the resulting models perform quite well on benign data, they are significantly more robust against targeted adversarial attacks: Our final, proposed model shows a performance on clean data similar to the baseline model, while being more than four times more robust.


\end{abstract}

\section{Introduction}

    ASR systems are ever more widely used in our daily life. They are being easily incorporated in our phones or in home electronics, which are not only used for human-machine interaction, but are also able to control other smart devices. Previous works have shown the vulnerability of these systems by adversarial attacks in the audio domain \cite{Lea_1, pmlr-v97-qin19a, Nicolas_1,Nicolas_2}, where malicious noise is added into the input signal to fool the ASR system and manipulate the predicted transcript. This opens the doors for attackers to breach the system security and violate users' privacy, \cite{Lea_1} pointed out in the introduction some real-world impact, such as unwanted online shopping orders made by amazon's devices.

    Many of the most widely used ASR systems are based on hybrid models, combining a deep neural network (DNN) with a hidden Markov model (HMM). The DNN predicts the class probabilities per speech frame, while the HMM is used to describe the temporal, context-dependent information that is given by the pronunciation, the vocabulary, and grammar of the language. Research has centered on the properties that make DNNs vulnerable to adversarial attacks not only in image recognition \cite{Christian_1, andrew_1}, but also in speech recognition \cite{pmlr-v97-qin19a, lea_2}, where specific perturbations are added to the input signal.
    These perturbations are chosen in such a way that they are hardly perceivable for humans but maximize the loss function of the model and lead the system to predict the output desired by the attacker \cite{Nicolas_3,Anug_1,jernej_1}. 

     The human perception of audio signals depends among other things on the frequency. The presence of one sound can render another sound imperceptible; this effect is known as masking. This happens most easily, when the two sounds are in the same so-termed critical band. These bands can be obtained from an auditory filterbank model that describes the behavior of the Cochlea, as analyzed in psychoacoustics. High frequencies have wider critical bandwidths than lower frequencies, which makes two sounds at high frequencies less distinguishable for humans and especially promising for adversarial attackers. Prior works have already explored this property, creating adversarial perturbations that are mostly hidden from human perception, raising the question of how to secure our current ASR systems are under attack \cite{Lea_1, lea_2}.
    
    There has thus been increasing interest in the research community in building robust ASR systems. However, there are only a few studies that tackle this problem and propose defenses against malicious audio \cite{waveguard,nicholas_3} and image perturbations \cite{battista_1,aleksander_1}. The ones that exist lack robustness against adaptive adversaries in white box settings, where the architecture of the model is known by the attacker. 

    Motivated by the observation that adversarial perturbations often lie in the high-frequency range, we propose to remove faster variations of the input signal. To do so, we use slow feature analysis (SFA)  \cite{Laurenz_2,6790128}, an unsupervised learning algorithm that is able to extract slow components of an input signal that varies on a faster time scale. It has been successfully applied in other fields, e.g., for extracting driving forces \cite{Laurenz_1}, nonlinear blind source separation \cite{inproceedings}, and for automatic lipreading \cite{Freiwald_1}. 
    In addition to SFA, we apply a low-pass filter with a cutoff frequency of 7 kHz\footnote{Baer et al.~have shown improvements on the intelligibility of speech in noise for people without dead regions in high frequencies by using low-pass filters, increasing the cutoff frequency up to 7.5 kHz \cite{Baer}.}. 
    Thus, we limit the possible space in which malicious noise can affect the recognition system. 
    We propose to apply these modifications to the audio signals before feeding them to hybrid DNN-HMM ASR systems (for training as well as for testing), 
    and present an empirical analysis that demonstrates the success of this strategy.
    While the performance on vanilla audio signals is not affected strongly, the rate of successful targeted adversarial attacks is reduced significantly.
  
    
    In Section \ref{section:back} we give a brief overview of the required background knowledge, in Section \ref{section:appr}, we describe our defense approach in more detail, and in Section \ref{section:exp}, we present our empirical analysis.
    
    \section{Background}
    \label{section:back}

    \subsection{Adversarial examples}

    For simplifying descriptions, we assume that the relationship between the  input audio signal \(x\) and the label transcript \(y\) is given by 
    \begin{equation*}
      y = f(x) \enspace,
    \end{equation*}
    where \(f(\cdot)\) is the mapping function of the ASR system that maps an input to the sequence of words it most likely corresponds to. The goal of a targeted adversarial attack is to add a minimal perturbation $\delta$ onto \(x\) such that the system predicts a new sequence of words \(\hat{y}\) defined by an attacker, i.e.
    \begin{equation*}
      f(x + \delta ) =\hat{y} \neq y = f(x) \enspace.
    \end{equation*}
    To generate an adversarial example, first, given an audio signal and its corresponding malicious text, we need to find an optimal time-aligned version, and second, we need to add the perturbation $\delta$ 
    that leads the
    ASR system to predict the desired transcript. 
    
    The first task is achieved by applying the forced alignment algorithm, which---given the original audio signal and the target transcription---generates an optimally time-aligned state sequence that matches the text with the audio signal. For the second task, we use the projected gradient descent (PGD) method for creating adversarial examples \cite{aleksander_1}. For an in-depth description of the generation of adversarial samples for ASR systems, we refer the reader to \cite{Lea_1}. PGD corresponds to solving
    a constrained optimization problem by maximizing the loss function of the network through gradient descent with respect to \(\delta \). In order to control the level of noise, the parameter \(\delta \) is constrained by \(\epsilon\), such that \(\|\delta\| \leq \epsilon \).
    
    \subsection{Slow feature analysis}
    Slow feature analysis is an unsupervised machine learning algorithm that can be used for dimensionality reduction or for extracting slow components from temporally correlated data. 
    
    Given an $N$-dimensional input signal  \(x(t)\) that varies through time, the goal is to find a nonlinear function \(g\)
    \begin{equation}
      y(t) = g(x(t))  \enspace,
      \label{eq3}
    \end{equation}
    such that the components \(y_i(t), i \in [1, ..., M]\), (sorted by the degree of slowness) of the $M$-dimensional output signal $y(t)$ minimize 
    \begin{equation*}
      \Delta(y_i) := \langle \Dot{y}_i^2 \rangle_t \enspace,
    \end{equation*}
    where \(\langle \cdot \rangle_t\) denotes the temporal average and \(\Dot{y}_i^2\) the square derivative of the output features. The set of outputs should carry as much information as possible, avoiding a trivial constant output signal. We achieve this by constraining the optimization problem as follows
    \begin{equation}
      \forall i: \langle {y_i} \rangle_t = 0
      \label{eq5} \enspace,
    \end{equation}
    \begin{equation*}
      \forall i: \langle {y_i^2} \rangle_t = 1 \enspace,
    \end{equation*}
    \begin{equation*}
      \forall i < j :\langle y_i y_j \rangle_t = 0 \enspace.
    \end{equation*}
    Hence, we remove the mean, stretch the signal to have unit variance in all directions, and decorrelate the output components. 

    \subsection{Low-pass filter}
    ASR systems are often trained with a sampling frequency of 16 kHz, representing the frequency content of the input signal between 0-8000 Hz. A low-pass filter attenuates signal components above a given cutoff frequency. Thus, by applying a low pass filter we restrict the effective space where an attacker can incorporate noise.
    
    \section{Approach}
    \label{section:appr}

    The goal of our defense approach is to make the hybrid ASR system more robust against targeted attacks by relying on slower components of its input signal. That is, we aim at making it more difficult for the attacker to
    manipulate the input in such a way that the ASR system produces a desired output. Towards this aim, we augment the ASR system by adding a pre-processing
    stage. In this stage, the input signal is transformed by applying SFA, a low pass filter, or both. Thus, our acoustic model is trained and tested using a modified version of the original data set. In the following, we describe the pre-processing steps in more detail:
    
    \begin{figure*}
    \captionsetup{font=footnotesize, skip=0pt}
    \begin{multicols}{2}
        \includegraphics[width=\linewidth]{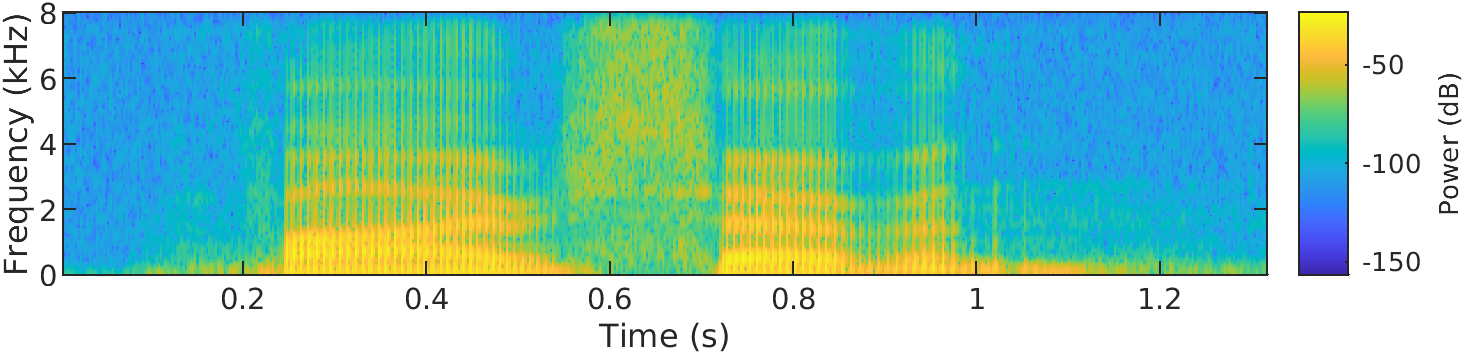}\par 
        \caption*{(a) Original audio signal power spectrum with transcription ``Five Seven"}
        \includegraphics[width=\linewidth]{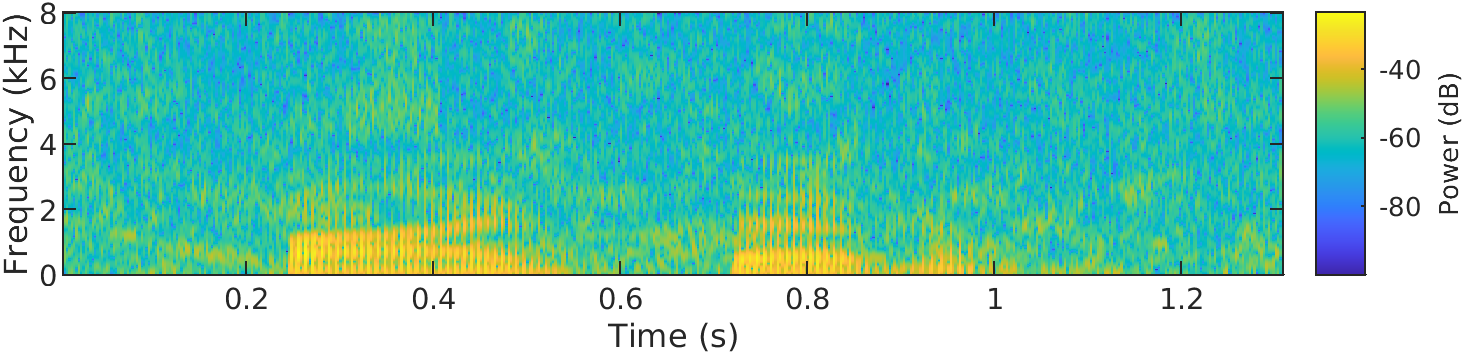}\par 
        \caption*{(b) Adversarial audio signal power spectrum with transcription ``Eight Six Five"}
        \end{multicols}
    \begin{multicols}{2}
        \includegraphics[width=\linewidth]{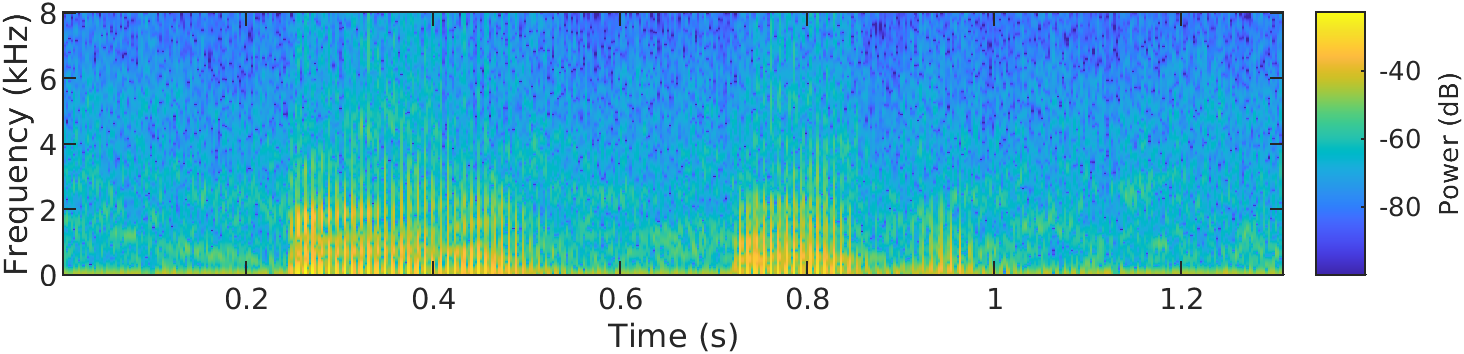}\par
        \caption*{(c) Adversarial audio signal transformed by SFA with model-predicted transcription ``Five Five"}
        \includegraphics[width=\linewidth]{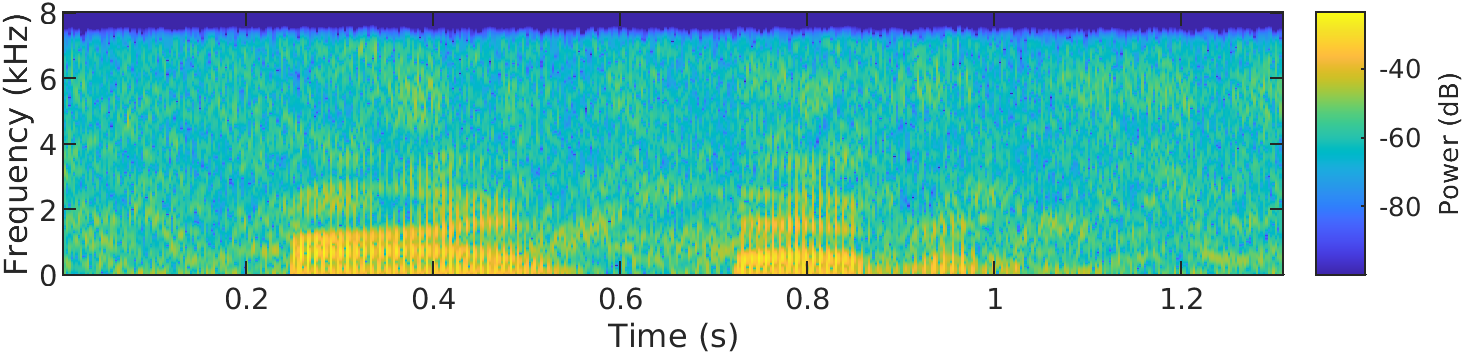}\par
        \caption*{(d) Adversarial audio signal after low-pass filtering, with model-predicted transcription ``Five Five"}
    \end{multicols}
    \caption{Spectrogram of original audio sample (1a) in comparison to the spectrogram of a corresponding adversarial audio sample before (1b) and after applying SFA (1c) or a low-pass filter (1d).}
    \label{fig:spect}
    \end{figure*}

    \subsection{Performing slow feature analysis}
    To extract the slowest component from 
    an input signal,
    we use the python data processing framework MDP \cite{neuro} to complete the following steps:
    \\[3pt]
    \noindent \textbf{Step 1: Time embedding.} We create a new multivariate time series by framing the input signal \(x(t)\) using a shifted delay of \(L\) elements. Thus, we expand the dimensionality of our data
    
    \begin{equation*}
      x_t(k, \tau) = x_t(k + R\tau), \quad k = 0,...,L-1,
    \end{equation*}
    
    \noindent where \(k\) represents the sample index, \(\tau\) the frame index, \(R\) the length of overlapping elements and \(L\) the length of the window. In our work \(L=2\) and \(R=1\).
    \\[3pt]    
    \noindent \textbf{Step 2: Non-linear expansion.} We expand our framing result from Step 1 by a quadratic function \(h\) to create non-linear features of our original data. 
    
    \begin{equation*}
      h(x) = [x_0, x_1, x_0^2, x_0x_1, x_1^2]^T \enspace.
    \end{equation*}

    \noindent \textbf{Step 3: Whitening.} To meet the constraints of the optimization problem in Equation \eqref{eq5}, the expanded signal is whitened. This removes the mean and stretches the signal along its principal axes, such that it has zero mean and unit variance in all directions. Moreover, by projecting the whitened signal onto two orthogonal unit vectors, the projected signal components are decorrelated. We use principal component analysis to sphere our data from Step 2. 
    \begin{equation*}
      z(t) := W(h(x(t))),
    \end{equation*}
    where \(W\) denotes the whitening function \cite{Thameri}.
    \\[3pt]
    \noindent \noindent \textbf{Step 4: Derivative.} We calculate the time derivative denoted by \(\Dot{z}_i(t)\), which, in our problem, is simply the difference between two successive time points.
    
    \begin{equation*}
      \Dot{z}_i(t) = z_i(t+1) - z_i(t).
    \end{equation*}
    
    \noindent \textbf{Step 5: Input-output function.} In order to find the direction, in which the signal varies most slowly, we first calculate the covariance matrix of our derivative signal from Step 4. 
    
    \begin{equation*}
      cov(\Dot{z}(t)) = \langle \Dot{z}\Dot{z}^T \rangle_t
    \end{equation*}
    
    \noindent Second, we perform a principal component analysis to find the smallest eigenvalue \(\lambda\), with its corresponding eigenvector \(w\) by solving the eigenvalue equation given by
    
    \begin{equation*}
        \langle \Dot{z}\Dot{z}^T \rangle_t w_j = \lambda_j w_j
              \quad \mbox{ with } \lambda_1 \leq \lambda_2 \leq ... \lambda_j \enspace.
      \label{eq11}
    \end{equation*}

    \noindent Finally, we choose \(\lambda_1\) and its corresponding \(w_1\), which corresponds to the direction of slowest variation, to do the projection of our data.
    \\[3pt]
    \noindent If we were interested in obtaining additional slow components, we could simply take the next, larger
    eigenvalue and project our data using its corresponding eigenvector. Notice that \(\lambda_j\) represents the slowness of the features, i.e.~the smaller the \(\lambda_j\), the slower a \(y_j\) we get. In our work, we extract only the slowest component given by \(\lambda_1\) and \(w_1\).
    \\[3pt]
    \noindent Finally, the input-output function from Equation \eqref{eq3} is determined just by a matrix multiplication:
    \begin{equation*}
        y_j(t) = g_j(x(t)) = w_j^T h(x(t)),
    \end{equation*}
    where \(w_j\) are the transformation vectors derived above. 
    
    \subsection{Low-pass filtering}
    We optionally apply a low-pass filter with a cutoff frequency of 7 kHz and a stop band frequency of 7.5 kHz, removing the high-frequency components of the input signal \cite{Monson_1}. 
    
    \section{Experiments}
    \label{section:exp}

    In the following, we describe our experimental analysis.

    \subsection{Speech recognition system}
    \label{subsection:ASR}

    We work in a PyTorch environment and train a hybrid DNN-HMM ASR system on the small-vocabulary TIDIGITS data set, which contains approximately 8\,000 utterances
    with a vocabulary consisting of digits from 0 to 9, with 2 alternative pronunciations for 0 (i.e.~'oh' and 'zero'). 
    
    The NN architecture has two hidden layers with 100 units each; the activation functions of the hidden layers are ReLUs, while the activation function of the output layer is a softmax function with 95 units that correspond to the states of the HMM. The input layer size (39) is chosen to work with feature vectors containing the first 13 mel-frequency cepstrum coefficients alongside their first and second derivatives. We employ 
    Adam  with an initial learning rate of 0.0001 \cite {kingma2017adam} as the optimizer of the network and the cross-entropy loss function. To improve results, after training the NN for 3 epochs, 5 
    epochs of Viterbi training are executed \cite{Jelinek}.

    In order to find the pre-processing leading to the best defense against adversarial attacks, we train five ASR systems, all of them with the same NN architecture described above, but after applying different kinds of pre-processing to the training set. 
    That is we perform SFA, low-pass filtering, or both techniques, as described in Section \ref{section:appr}.
    We repeat the experiments five 
    times with different random initializations 
    and
    report the mean performance. 
    An example of the spectrogram of one input signal, the corresponding adversarial example, and the same adversarial example after pre-processing are shown in Figure \ref{fig:spect}.

    Overall, we use these models in the subsequent investigations:
    
    \begin{enumerate}
        \item \textit{Baseline}: NN trained with original training data;
        \item \textit{Bas\_SFA}: NN trained with original training data and data pre-processed by SFA;  
        \item \textit{SFA}: NN trained only with data pre-processed by SFA; 
        \item \textit{LPF}: NN trained only with data pre-processed with a low pass filter;  
        \item \textit{SFA\_LPF}: NN trained with data pre-processed by SFA
        and the low pass filter. 
    \end{enumerate}
    
    \subsection{Adversarial attack}
    To generate the target transcriptions of the adversarial examples, digits
    from  0 to 9 are randomly selected. The total number of digits per target transcript varies randomly between 1 and 5. To find the optimally time-aligned state sequence denoted by \({x}^{*}\) for a given combination of an audio signal and an adversarial text, we use the Montreal forced aligner \cite{McAuliffe}. To generate the adversarial audio signal that generates this desired \({x}^{*}\), we use PGD as provided by the library Cleverhans \cite{papernot2018technical}. 

    We refer the reader to the description in the work of Däubener et al.~\cite{sina_1} for more details on the creation of the adversarial examples.
   
    \subsection{Model evaluation}
    To evaluate the performance of the speech recognition models, we use the word error rate (WER) given by 
    
    \begin{equation}
         \text{WER} = 100 \cdot \frac{S+D+I}{N} \enspace,
      \label{eq13}
    \end{equation}
    where \(S\) is the number of substituted words, \(D\) is the number of deleted words and \(I\) is the number of inserted words. \(N\) is the total number of words in the reference text, which corresponds to the ground-truth label of the original test sample or the malicious target transcription for the adversarial attack, respectively.
    We aim at a model that has a low WER on the original data, but a high WER compared to the target transcript of an adversarial attack, since this means that the attack is not successful.
     
    The WER is evaluated on a test set of 1\,000 malicious samples and 1\,000 original test samples.
 
     \begin{table}[th]
     \footnotesize
      \captionsetup{font=footnotesize, skip=0pt}
      \caption{Average WER 
      of different models for 1\,000 test samples  and adversarial examples created with $\epsilon=0.5$.
      }
      \label{tab:wer_no_tra}
      \centering
      \begin{tabular}{ l | c c c| c c c }
        \toprule
        \multicolumn{1}{c|}{} &
        \multicolumn{3}{c|}{\textbf{Original Test set}} &
        \multicolumn{3}{c}{\textbf{Adversarial Test set}} \\
        & WER & Std Dev & W & WER & Std Dev & W \\
        \midrule
        Baseline & 0.79\% & 0.0013 & - & 10.46 \% & 0.0252 & - \\ 
        Bas\_SFA & 1.01\% & 0.0009 & 16 & 11.86\% & 0.0159 & 21 \\  
        SFA & 2.71\% & 0.0071 & 15 & 12.66\% & 0.0183 & 21 \\ 
        LPF & 1.07\% & 0.0012 & 16.5 & 11.78\% & 0.0156 & 23 \\ 
        SFA\_LPF & 2.62\% & 0.0058 & 15 & 9.99\% & 0.0118 & 27 \\           
        \bottomrule
      \end{tabular}
    \end{table}
    
    \subsection{Results}



    

    The average WERs of the different models for the 1\,000 samples from the test set and the 1\,000 adversarial samples are shown in Table~\ref{tab:wer_no_tra}. We observe a low WER of 0.79\% for the baseline model on the original test set. For the adversarial test set, all models are seen to be unprotected against adversarial perturbations.

    When the same pre-processing step that is applied during training\footnote{We did not include pre-processing during training for baseline, as the NN was not trained under any data transformation condition.} is also used in the prediction procedure during inference, the performance on the original test set only varies slightly, while the WER for adversarial examples increases significantly, see Table \ref{tab:wer_tra}.  The LPF model is especially promising, since it performs similar to the baseline on the original test set, but is significantly more ros. In the first phase, input data transformation bust to adversarial attacks.   The largest improvement in robustness is given by the SFA\_LPF model,  but at the cost of aggravating the WER on clean data.
    

   To claim that a model performs significantly  different to the baseline model 
   the Wilcoxon Rank-Sum Test (WRST) W statistic is provided in both tables. Enough statistical evidence for the corresponding  claim exists when the W statistic is less than 17.\footnote{The lower critical value for the WRST for independent samples with sizes of 5 is 17, corresponding to $\alpha$  = 0.05 for a two-tailed test.}

    \begin{table}[th]
    \footnotesize
    \captionsetup{font=footnotesize, skip=0pt}
      \caption{Average WER 
      of models applied to pre-processed data (with different kind of data transformations) for 1\,000 test samples  and adversarial examples created with $\epsilon=0.5$
      Most promising results are printed bold. }
      \label{tab:wer_tra}
      \centering
      \begin{tabular}{ l | c c c | c c c }
        \toprule
        \multicolumn{1}{c|}{} &
        \multicolumn{3}{c|}{\textbf{Original test set}} &
        \multicolumn{3}{c}{\textbf{Adversarial test set}} \\
        & WER & Std Dev & W & WER & Std Dev & W\\
        \midrule
        Baseline & 0.79\% & 0.0013 & - & 10.46\% & 0.0252 & - \\ 
        Bas\_SFA & 2.76\% & 0.0018 & 15 & 48.01\% & 0.0364 & 15 \\  
        SFA & 2.28\% & 0.0012 & 15 & 43.50\% & 0.0704 & 15  \\ 
        LPF & \textbf{0.80}\% & 0.0014 & 27.5 & \textbf{44.32}\% & 0.0267 & 15 \\ 
        SFA\_LPF & \textbf{2.21}\% & 0.0024 & 15 & \textbf{94.14\%} & 0.0367 & 15 \\           
        \bottomrule
      \end{tabular}
    \end{table}

\begin{figure}[t]
    \captionsetup{font=footnotesize, skip=0pt}
      \includegraphics[width=\linewidth]{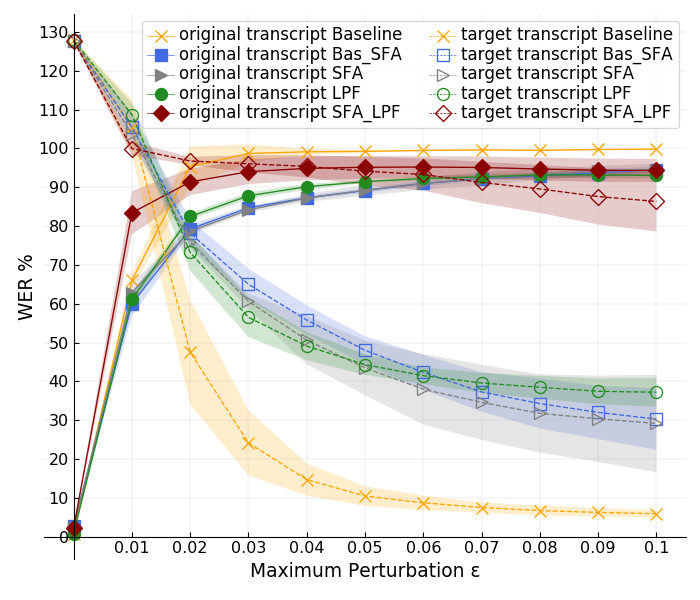}
      \caption{Mean WER and standard deviation over 5 runs, comparing the predicted transcript and the original text or the adversarial text for different models and perturbation levels \(\epsilon\).}
      \label{fig:wer_prep}  
    \end{figure}

    Figure \ref{fig:wer_prep} illustrates the effect of adversarial noise at perturbation levels \(\epsilon \in [0, 0.1]\) with increments of 0.01. For \(\epsilon = 0\), the sample corresponds to the original test point. 
    For calculating the WERs, we again created 1\,000 adversarial samples corresponding to 1\,000 test samples per model and per perturbation level \(\epsilon\) for each run. 
    
    Although all analysed pre-processing strategies decreased the success rate of adversarial attacks (which improves when increasing the level of perturbation \(\epsilon \)), it comes at the cost of harming the performance on clean data. 
    Only the model combined with the LPF has no significant performance decrease, while exhibiting a WER four times larger than the baseline model for adversarial examples.
    When solely considering robustness, the best strategy is to combine SFA and the low-pass filter, 
    yielding
    an average WER above 80\% 
    for all levels of perturbation. 
    The results  support our hypothesis that removing high frequency components and relying on the slowest features limits the space, where adversarial noise can be added effectively, 
    which in turn makes the ASR less vulnerable against adversarial attacks.
    
    \section{Discussion }
    In training the different ASR models, we have implemented a set of potential data transformations, and used this data to train small-scale acoustic models from scratch. It would be interesting to investigate whether we can reduce the time-consuming part of training by reusing the lower layers of a pre-trained network, i.e.~whether our experiments can be improved by incorporating the principle of transfer learning. Computational effort is not a significant issue in the presented small-vocabulary experiments, but will be interesting when running our experiments for large-vocabulary tasks. We leave this implementation as an interesting direction for future work.
    
    \section{Conclusions}s. In the first phase, input data transformation 
    
    In this work, we show that extracting slow variations and removing high-frequency components of audio signals that serve as input to an ASR can make the system more robust against targeted adversarial attacks.
    When these data transformations were added as pre-processing steps during training and inference, the success rate of targeted adversarial examples was decreased significantly. For most data transformations, this came at a cost of 
    significantly decreased
    performance on clean data. 
    However the LPF did not lead to a performance decrease on clean data, while resulting in a model significantly more robust than the baseline.
    A combination of SFA and LPF induced the highest robustness 
    but also more the doubles the WER on clean data.

    Of course our results are based on a limited experimental analysis. Future research will 
    test the transferability of our findings to large-vocabulary ASR systems and the creation of adaptive attacks to evaluate the defense against a skilled attacker. 
    
    \section{Acknowledgements}

    This work was supported by the German Academic Exchange Service - 57451854, the Chilean government (CONICYT-PFCHA/Doctorado  acuerdo bilateral DAAD) - 62180013 and partially funded by the Deutsche Forschungsgemeinschaft (DFG, German Research Foundation) under Germany's Excellence Strategy  - EXC 2092 CASA – 390781972.

    \bibliographystyle{IEEEtran}
    
    \bibliography{mybib}
    
\end{document}